\documentclass[twocolumn,10pt]{IEEEtran_v15} 
\usepackage{mathsym}
\usepackage{amsmath}
\usepackage{amssymb}
\usepackage{epsf}
\usepackage{latexsym}
\usepackage[dvips]{graphicx}
\renewcommand{\baselinestretch}{0.92}

\def\BibTeX{{\rmfamily B\kern-.05em{\scshape i\kern-.025em b}\kern-.08em \TeX}}

\begin{document}

\title{\bf \LARGE Trellis-Based Equalization for Sparse ISI Channels Revisited}

\author{
\large Jan Mietzner$^1$,~ Sabah Badri-Hoeher$^1$,~ Ingmar Land$^2$,~ and~ Peter A. Hoeher$^1$\\[2ex]
\normalsize $^1$Information and Coding Theory Lab, Faculty of Engineering, University of Kiel, Kaiserstrasse 2, D-24143 Kiel, Germany.\\
\normalsize WWW: $\,$ {\tt http://www-ict.tf.uni-kiel.de} $\;\;$ E-mail: $\,$ {\tt \{jm,sbh,ph\}@tf.uni-kiel.de}\\[1ex]
\normalsize $^2$Digital Communications Division, Department of Communication Technology, Aalborg University, Fredrik Bajers Vej 7, A3,\\ \normalsize  9220 Aalborg {\O}, Denmark. $\;\;$$\;\;$\normalsize WWW: $\,$ {\tt http://typo3.kom.aau.dk/infsig/} $\;\;$ E-mail: $\,$ {\tt il@kom.aau.dk}
\vspace*{-0.8cm}}

\markboth{IEEE GLOBECOM 2002}
{Murray and Balemi: Using the Document Class infocom.cls}

\maketitle
\pagestyle{empty}
\thispagestyle{empty}

\begin{abstract}
Sparse intersymbol-interference (ISI) channels are encountered in a variety of high-data-rate communication systems. Such channels have a large channel memory length, but only a small number of significant channel coefficients. In this paper, trellis-based equalization of sparse ISI channels is revisited. Due to the large channel memory length, the complexity of maximum-likelihood detection, e.g., by means of the Viterbi algorithm (VA), is normally prohibitive. In the first part of the paper, a unified framework based on factor graphs is presented for complexity reduction without loss of optimality. In this new context, two known reduced-complexity algorithms for sparse ISI channels are recapitulated: The multi-trellis VA ($\mbox{M-VA}$) and the parallel-trellis VA ($\mbox{P-VA}$). It is shown that the $\mbox{M-VA}$, although claimed, does not lead to a reduced computational complexity. The $\mbox{P-VA}$, on the other hand, leads to a significant complexity reduction, but can only be applied for a certain class of sparse channels. In the second part of the paper, a unified approach is investigated to tackle general sparse channels: It is shown that the use of a linear filter at the receiver renders the application of standard reduced-state trellis-based equalizer algorithms feasible, without significant loss of optimality. Numerical results verify the efficiency of the proposed receiver structure. 
\end{abstract}

\begin{keywords}
Trellis-based equalization, sparse ISI channels, complexity reduction, prefiltering.
\end{keywords}\vspace{-0.1cm}

\section{Introduction}\label{intro}
\PARstart{S}{parse} intersymbol-interference (ISI) channels are encountered in a wide range of communication systems, such as high-data-rate mobile radio systems (especially in hilly terrain), wireline systems, or aeronautical/ satellite communication systems. For mobile radio applications, fading channels are of particular interest. The equivalent discrete-time channel impulse response (CIR) of a sparse ISI channel has a large channel memory length $L$, but only a small number of significant channel coefficients.

Due to the large memory length, equalization of sparse ISI channels is a demanding task. The topics of linear equalization and decision-feedback equalization for sparse ISI channels are, e.g., addressed in \cite{LE_DFE}, where the sparse structure of the channel is explicitly utilized for the design of the corresponding finite-impulse-response (FIR) filter(s). Trellis-based equalization for sparse channels is addressed in \cite{trellis1}-\cite{parallelMAP}. The complexity in terms of trellis states of an optimal trellis-based equalizer, based on the Viterbi algorithm (VA)~\cite{viterbi} or the Bahl-Cocke-Jelinek-Raviv algorithm (BCJRA)~\cite{bcjr}, is normally prohibitive for sparse ISI channels, because it grows exponentially with the channel memory length $L$\footnote{The VA is optimal in the sense of maximum-likelihood sequence estimation (MLSE) and the BCJRA in the sense of maximum a-posteriori (MAP) symbol-by-symbol estimation. Both algorithms operate on the same trellis diagram. All statements concerning complexity hold for both the VA and the BCJRA.}. However, reduced-complexity algorithms can be derived by exploiting the sparseness of the channel. 

In \cite{trellis1}, it is observed that given a sparse channel, there is only a comparably small number of possible branch metrics within each trellis segment. By avoiding to compute the same branch metric several times, the computational complexity is reduced significantly without loss of optimality. However, the complexity in terms of trellis states remains the same and thus the storage expense. As an alternative, another equalizer concept coined {\sl multi-trellis Viterbi algorithm ($\mbox{M-VA}$)} is proposed in \cite{trellis1} that is based on multiple parallel {\it irregular} trellises (i.e., time-variant trellises). The $\mbox{M-VA}$ is claimed to have a significantly reduced computational complexity and storage expense without (much) loss of optimality. 

A particularly simple solution to reduce the complexity of the conventional VA without loss of optimality can be found in~\cite{trellis3}: The {\sl parallel-trellis Viterbi algorithm ($\mbox{P-VA}$)} is based on multiple parallel {\it regular} trellises. However, an application of the $\mbox{P-VA}$ is only possible for a certain class of sparse channels having a so-called {\sl zero-pad structure}. In order to tackle more general sparse channels with a CIR close to a zero-pad channel, it is proposed in \cite{trellis3} to exchange tentative decisions between the parallel trellises and thus cancel residual ISI. This modified version of the $\mbox{P-VA}$ is, however, suboptimal and is denoted as $\mbox{\sl \underline{sub}-P-VA}$ in the sequel. 

A generalization of the $\mbox{P-VA}$ and the $\mbox{sub-P-VA}$ can be found in~\cite{parallelMAP}, where corresponding algorithms based on the BCJRA are presented. These are in the sequel denoted as {\sl parallel-trellis BCJR algorithms} ($\mbox{\sl P-BCJRA}$ and $\mbox{\sl sub-P-BCJRA}$, respectively). Some interesting enhancements of the $\mbox{(sub-)P-BCJRA}$ are also discussed in \cite{parallelMAP}. Specifically, it is shown that the performance of the $\mbox{sub-P-BCJRA}$ can be improved by means of minimum-phase prefiltering \cite{prefilter,SabahPeter_Kalman} at the receiver. A specific FIR approximation of the infinite-length linear minimum-phase filter is used, which preserves the sparse structure of the channel. This guarantees that the $\mbox{sub-P-BCJRA}$ can still be applied after the prefiltering. 

Alternatives to trellis-based equalization are the tree-based LISS algorithm \cite{liss} and the Joint Gaussian (JG) approach in~\cite{JG}. In this paper, trellis-based equalization for sparse ISI channels is revisited. A unified framework based on factor graphs~\cite{factorgraph} is presented in Section~\ref{framework} for complexity reduction without loss of optimality, and the $\mbox{M-VA}$ as well as the $\mbox{P-VA}$ are recapitulated in this new context. Specifically, it is shown that the $\mbox{M-VA}$ does, in fact, {\it  not} lead to a reduction of computational complexity, compared to the conventional VA. Moreover it is illustrated, why the optimal $\mbox{P-VA}$ can only be applied for zero-pad channels. In order to equalize general sparse ISI channels, a simple alternative to the $\mbox{sub-P-VA}$/ $\mbox{sub-P-BCJRA}$ is investigated in Section~\ref{prefiltering}. For this purpose, the idea in \cite{parallelMAP} to employ prefiltering at the receiver is picked up.  It is demonstrated that the use of a linear minimum-phase filter renders the application of reduced-state equalizers such as \cite{reducedstate,ddfse} feasible, without significant loss of optimality. The proposed receiver structure is notably simple: The employed equalizer algorithms are standard, i.e., not specifically designed for sparse channel, because the sparse channel structure is normally lost after prefiltering. Solely the linear filter is adjusted to the current CIR (which is particularly favorable with regard to fading channels), where the filter coefficients can be computed according to standard techniques available in the literature  \cite{prefilter,SabahPeter_Kalman}. In order to illustrate the efficiency of the proposed receiver structure, numerical results are presented for various types of sparse ISI channels.  Bit error rates (BERs) are achieved that deviate only 1-2~dB from the matched filter bound (at a BER of $10^{-3}$). To the authors' best knowledge, similar performance studies for prefiltering in the case of sparse ISI channels have not yet been presented in the literature.  

\section{Complexity Reduction without Loss of Optimality}\label{framework}
A {\sl general sparse ISI channel} has a comparably large channel memory length $L$, but only a small number of significant channel coefficients $h_g$,~ \mbox{$g = 0,...,G \ll L$}, according to
\begin{equation}\label{sparseCIR}
{\mathbf h} \;:=\; [\, h_0\; \underbrace{0 \hdots 0}_{ f_0 \; {\rm zeros}}\; h_1 \; \underbrace{0 \hdots 0}_{ f_1\; {\rm zeros}}\; \;\hdots \;\;\underbrace{0 \hdots 0}_{ f_{G-1}\; {\rm zeros}} h_G \,]^{\rm T}\,, 
\end{equation}
with $f_i\!\geq\!0$ integer for all $i$ and \mbox{$L = \sum_{i=0}^{G-1} (f_i + 1)$}. A sparse ISI channel, for which $f_0 = f_1 = ... = f_{G-1} =: f$ holds, is referred to as {\sl zero-pad channel} \cite{trellis3}. 

Throughout this paper, complex baseband notation is used. The $k$-th transmitted $M$-ary data symbol is denoted as $x[k]$, where $k$ is the time index. A hypothesis for $x[k]$ is denoted by $\tilde{x}[k]$ and a hard decision by $\hat{x}[k]$. For simplicity, the channel coefficients are assumed to be constant over an entire block of data symbols (block length $N\! >\! L$). The equivalent discrete-time channel model is given by 
\begin{equation}\label{ISIchannelmodel}
y[k] \;=\; h_0 \, x[k] \;+\; \sum_{g=1}^{G} \, h_g \, x[k-d_g] \;+\; n[k]\,,
\end{equation}
where $y[k]$ denotes the $k$-th received sample and $n[k]$ the $k$-th sample of a complex additive white Gaussian noise (AWGN) process with zero mean and variance $\sigma^2_{\rm n}$. Moreover, 
\begin{equation}
d_g \;:=\; \sum_{i=1}^{g} (f_{i-1} + 1), \hspace{0.5cm} 1 \leq g \leq G,
\end{equation}
denotes the position of $h_g$ in ${\mathbf h}$.

In the sequel, the channel vector ${\mathbf h}$ is assumed to be known at the receiver. Moreover, an $M$-ary alphabet for the data symbols is assumed. The complexity in terms of trellis states of the conventional Viterbi/ BCJR algorithm is given by ${\cal O}\{M^L\}$ and is thus normally prohibitive. Given a zero-pad channel, it is proposed in \cite{trellis3} to decompose the conventional trellis diagram with  $M^L\! =\! M^{(f+1)G}$ states into $(f\!+\!1)$ parallel regular trellises, each having only $M^G$ states. As will be shown in the sequel, a decomposition into multiple parallel regular trellises is {\it not} possible in the case of a more general sparse ISI channel. 

In order to decompose a given trellis diagram into multiple parallel trellises, the following question is of central interest: Which symbol decisions $\hat{x}[k]$ are influenced by a certain symbol hypothesis $\tilde{x}[k_0]$? Suppose, a certain decision $\hat{x}[k_1]$ is {\it not} influenced by the hypothesis $\tilde{x}[k_0]$, where $k_0$ and $k_1$ are two arbitrary (but fixed) time indices. Furthermore, let the set $\hat{\cal X}_{k_0} \!:=\!\left\{\hat{x}[k] \,|\, \hat{x}[k]\right.$ depends on $\left. \tilde{x}[k_0]\right\}$ contain all decisions influenced by $\tilde{x}[k_0]$ and the set $\hat{\cal X}_{k_1}$ all decisions influenced by $\tilde{x}[k_1]$. If these two sets are disjoint, i.e.~$\hat{\cal X}_{k_0}\cap\hat{\cal X}_{k_1}\!=\!\emptyset$, the hypotheses $\tilde{x}[k_0]$ and $\tilde{x}[k_1]$ can be accommodated in {\it separate} trellis diagrams without loss of optimality. In this case, a decomposition of the overall trellis diagram into (at least two) parallel regular trellises is possible.

\begin{figure*}[t]
\begin{center}
\includegraphics*[scale=0.54]{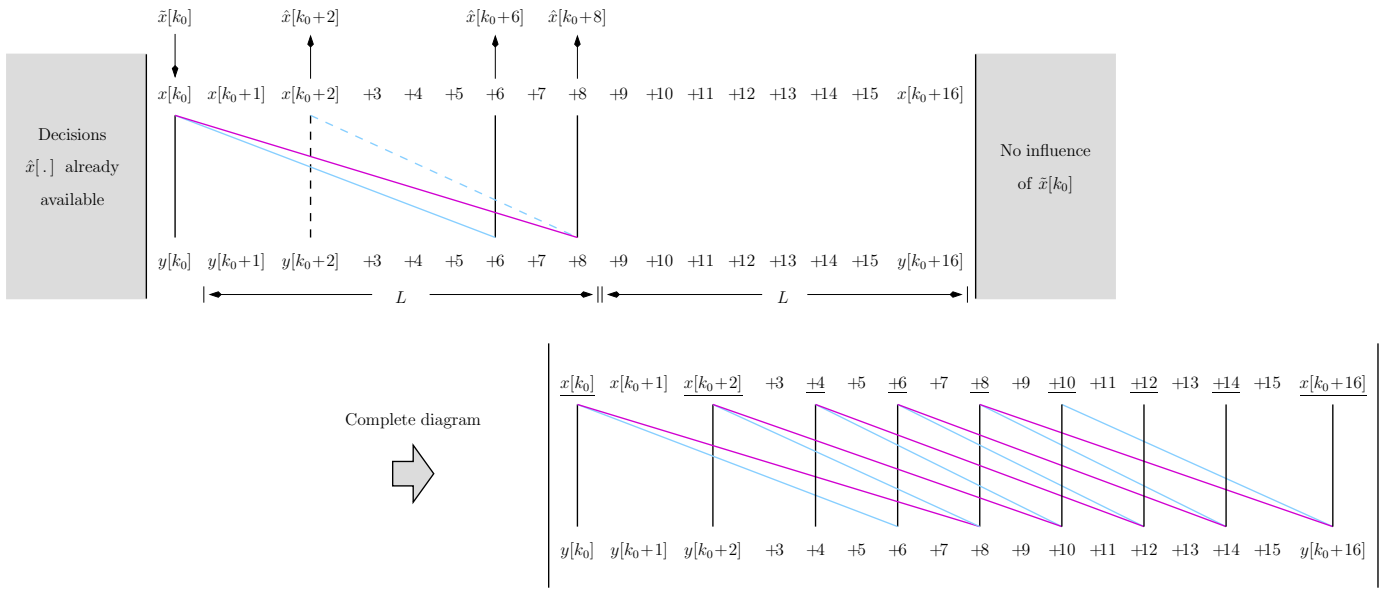} 
\end{center}\vspace*{-0.3cm}
\caption{Dependencies between symbol hypothesis $\tilde{x}[k_0]$ and subsequent decisions $\hat{x}[k]$: Example CIR~ ${\mathbf h}^{(1)} := [\, h_0\;\; 0 \;\; 0 \;\; 0 \;\; 0 \;\; 0 \;\; h_1 \;\; 0\;\;h_{2}\,]^{\rm T}$ ($L\!=\!8$, $G\!=\!2$).}
\end{figure*}

This fact is illustrated in the sequel for two example channels ($L\!=\!8$ and $G\!=\!2$ in both cases). The first channel is characterized by a CIR~ ${\mathbf h}^{(1)} := [\, h_0\; 0 \;\, 0 \;\, 0 \;\, 0 \;\, 0 \;\, h_1 \; 0\;\,h_{2}\,]^{\rm T}$ and the second by a CIR~ ${\mathbf h}^{(2)} := [\, h_0\; 0 \;\, 0 \;\, 0 \;\, 0 \;\, 0 \;\, 0\;\, h_1 \;h_{2}\,]^{\rm T}$. For the time being, a particular symbol hypothesis $\tilde{x}[k_0]$ is considered, and it is assumed that hard decisions $\hat{x}[k]$ are already available for all time indices $k\!<\!k_0$. Moreover, it is assumed that the hypothesis $\tilde{x}[k_0]$ does not have any impact on decisions $\hat{x}[k]$ with $k\!>\!k_0\!+\!DL$ ($D\!>\!0$ integer; in the example, we consider the case $D\!=\!2$)\footnote{This corresponds to the assumption that a VA with a decision delay of $DL$ symbol durations is optimal in the sense of MLSE.}. Fig.~1 shows a diagram for the first example CIR. The diagram may be interpreted as a factor graph~\cite{factorgraph} and illustrates the dependencies between hypothesis $\tilde{x}[k_0]$ and all decisions $\hat{x}[k]$, $k_0 \!\leq\! k \!\leq\! k_0\!+\!DL$. It can be seen from (\ref{ISIchannelmodel}) that only the received samples $y[k_0]$, $y[k_0\!+\!6]$, and $y[k_0\!+\!8]$ are directly influenced by the data symbol $x[k_0]$. Therefore, there is a direct dependency between the hypothesis $\tilde{x}[k_0]$ and the decisions $\hat{x}[k_0]$, $\hat{x}[k_0\!+\!6]$, and $\hat{x}[k_0\!+\!8]$. The received sample $y[k_0\!+\!8]$, for example, is also directly influenced by the data symbol $x[k_0\!+\!2]$. Correspondingly, there is as well a dependency between $\tilde{x}[k_0]$ and the decision $\hat{x}[k_0\!+\!2]$. The data symbols $x[k_0\!+\!6]$ and $x[k_0\!+\!8]$ again directly influence the received samples $y[k_0\!+\!12]$, $y[k_0\!+\!14]$, and $y[k_0\!+\!16]$, and so~on. 

Including all dependencies, one obtains the second graph of Fig.~1. As can be seen, there is a dependency between $\tilde{x}[k_0]$ and all decisions $\hat{x}[k_0\!+\!2\nu]$, where $\nu\!=\! 0,1,...,\lfloor DL/2 \rfloor$. Consequently, in this example it is possible to decompose the conventional trellis diagram into two parallel regular trellises, one comprising only the time indices $k_0\!+\!2\nu$ and the other one comprising the time indices $k_0\!+\!1\!+\!2\nu$. While the conventional trellis diagram has $M^8$ trellis states, there are only $M^4$ states in each of the two parallel trellises. (Moreover, a single trellis segment in the parallel trellises spans two consecutive time indices.) This result is in accordance with~\cite{trellis3}, since the CIR ${\mathbf h}^{(1)}$ constitutes a zero-pad channel $[\, h'_0\; 0 \; h'_1 \; 0 \; h'_2 \; 0 \; h'_3 \; 0\;h'_{4}\,]^{\rm T}$, where $G\!=\!4$, $f\!=\!1$,  and $h'_1\! =\! h'_2\! = \!0$. Generally spoken, a decomposition of a conventional trellis diagram into multiple parallel regular trellises is possible, if all non-zero channel coefficients of the sparse ISI channel are on a zero-pad grid with $f\!\geq\! 1$. In this case, the optimal $\mbox{P-VA}$ can be applied\footnote{The $\mbox{P-VA}$ is still optimal in the sense of MLSE. The finite decision delay $DL$ is not required and has only been introduced for illustrative purposes.}; otherwise one has to resort to the $\mbox{sub-P-VA}$ or to alternative solutions such as the $\mbox{M-VA}$. 

The second CIR ${\mathbf h}^{(2)}$ constitutes a counter example. Here, the symbol hypothesis $\tilde{x}[k_0]$ influences {\it all} decisions $\hat{x}[k']$, \mbox{$k_0 \!\leq \!k' \!\leq \!k_0\!+\!DL$} (not depicted due to space limitation). Consequently, a decomposition of the conventional trellis diagram into multiple parallel regular trellises is {\it not} possible here. Still, a decomposition into multiple parallel {\it irregular} trellises is possible, as proposed in \cite{trellis1} in the context of the $\mbox{M-VA}$. However, a significant subset of the dependencies resulting from the corresponding factor graph is neglected in \cite{trellis1}. If all dependencies are taken into account, the $\mbox{M-VA}$ does not yield any complexity advantage over the conventional VA.  The computational complexities in terms of the overall number of branch metrics that have to be computed for a single decision $\hat{x}[k_0]$ are stated in Table~1, for the conventional VA, the $\mbox{P-VA}$ (example CIR ${\mathbf h}^{(1)}$), and the $\mbox{M-VA}$ (example CIR ${\mathbf h}^{(2)}$). 

\section{Prefiltering for Sparse Channels}\label{prefiltering}
The preceding section has shown that trellis-based equalization of a general sparse ISI channel is quite a demanding task: An application of the optimal $\mbox{P-VA}$ (or the $\mbox{P-BCJRA}$) is only possible for zero-pad channels. In the case of a more general sparse channel, the suboptimal $\mbox{sub-P-VA}$ with residual ISI cancellation can be used. However, for a good performance the CIR should at least be close to a zero-pad structure \cite{trellis3}. The $\mbox{M-VA}$, on the other hand, was designed for sparse channels with a general structure, but does not offer any complexity advantage over the conventional VA if all dependencies between the individual symbol hypotheses are taken into account. 

\begin{table}[t]
\caption{Computational complexities:~ Viterbi algorithm~(VA), parallel-trellis VA ($\mbox{P-VA}$), and multi-trellis VA ($\mbox{M-VA}$).}
\begin{center}
{\small
\begin{tabular}{|c|c|c|}\hline 
{\bf Conventional VA}, & {\bf P-VA}, & {\bf M-VA}, \\
any CIR with $L\!=\!8$ & example CIR ${\mathbf h}^{(1)}$ & example CIR ${\mathbf h}^{(2)}$ \\ \hline \hline
  &  & ${\cal O}\{3\,M^9 + 2\,M^8$\\ 
 ${\cal O}\{M^9\}$&${\cal O}\{2\,M^5\}$ & $ +\; 2\,M^7 + 2\,M^6$ \\
 & & $  +\;2\,M^5 + 2\,M^4$\\
 & & $+\; 2\,M^3 + M^2\}$\\\hline
\end{tabular}}
\end{center}\vspace*{-2ex} 
\end{table}

In order to tackle general sparse ISI channels, a simple alternative to the $\mbox{sub-P-VA}$/ $\mbox{sub-P-BCJRA}$ is proposed in the sequel: It is demonstrated that the use of a linear minimum-phase filter at the receiver renders the application of standard reduced-state equalizer algorithms feasible, without significant loss of optimality. The receiver structure under consideration is illustrated in Fig.~2, where $z[k]$ denotes the $k$-th received sample after prefiltering and ${\bf h}_{\rm min}$ the filtered CIR. Within the scope of this paper, the ideal linear minimum-phase filter is approximated by an FIR filter of length $L_{\rm F}$, where the approach in \cite{SabahPeter_Kalman} is used to calculate the filter coefficients. The resulting FIR filter approximates a discrete-time whitened matched filter (WMF), i.e., the effect of noise coloring is negligible. The computational complexity of calculating the filter coefficients is ${\cal O}(L_{\rm F}\,L^2)$, i.e. only linear with respect to the filter length. Therefore, comparably large filter lengths are feasible. With regard to the trellis-based equalizer, we focus on delayed decision-feedback sequence estimation (DDFSE) \cite{ddfse} in the sequel. The number of trellis states in the DDFSE equalizer is $M^K$, where $K\!\ll\!L$ is a design parameter. In order to obtain a complexity that is similar to that of the (sub-)P-VA/ P-BCJRA equalizer, one should choose $K$ such that\footnote{In order to find an appropriate value for $K$ in the case of a general sparse ISI channel, one has to find an underlying zero-pad channel with a structure as close as possible to the CIR under consideration.}
\begin{equation}
K \;\leq\; {\rm log}_M (f+1) \;+\; G \,.
\end{equation}

In the following section, it is shown that the sparse channel structure is normally lost after prefiltering. Afterwards, numerical results are presented for various examples to demonstrate the efficiency of the proposed receiver structure.

\begin{figure}[t]
\begin{center}
\includegraphics*[scale=0.35]{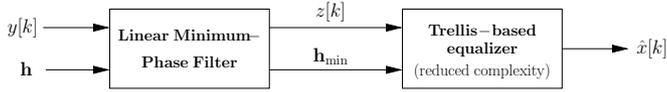} 
\end{center}
\caption{Receiver structure under consideration.}
\end{figure}

\subsection{Channel Structure After Prefiltering}\label{chstructure}
The sparse structure of a given CIR ${\bf h}$ is normally lost after minimum-phase prefiltering. For example, consider the CIR ${\bf h} = [\sqrt{0.5}\;\;0\;\;0\;\;\sqrt{0.1}\;\;\sqrt{0.4}]^{\rm T}$~ (i.e. $||{\bf h}||^2 = 1$), and let $H(z)$ denote the $z$-transform of ${\bf h}$. The zeros of $H(z)$ are given by
\begin{itemize}
\item $z_{0\;1,2}=0.69\,\pm\, {\rm j}0.80$~ ($|z_{0\;1,2}|=1.06$)~ and
\item $z_{0\;3,4}=-0.69\,\pm\, {\rm j}0.56$~ ($|z_{0\;3,4}|=0.89$).
\end{itemize}
Correspondingly, ${\bf h}$ is neither minimum-phase nor maximum-phase\footnote{Given a minimum-phase (maximum-phase) CIR ${\bf h}$, all zeros of $H(z)$ are inside (outside) the unit circle, i.e., $|z_{0,i}| \leq 1$ ($|z_{0,i}| \geq 1$) for all $i$.}. The $z$-transform $H_{\rm min}(z)$ of the filtered CIR  ${\bf h}_{\rm min}$ is obtained as follows: Those zeros of $H(z)$ that are inside or on the unit circle are retained for $H_{\rm min}(z)$, whereas the outside zeros are reflected into the unit circle. The resulting minimum-phase CIR is ${\bf h}_{\rm min}=[0.79\;\;0.12\;-0.02\;\;0.20\;\;0.56]^{\rm T}$, which is not sparse anymore.

As opposed to this, in the special case of a zero-pad channel the sparse structure is always preserved after minimum-phase prefiltering: Let ${\mathbf h} := [\, h_0\; h_1 \;\hdots \; h_G \,]^{\rm T}$ denote a (non-sparse) CIR with $z$-transform $H(z)$, and let ${\mathbf h}_{\rm ZP}$ denote the corresponding CIR with memory length $(f+1)G$ and $z$-transform $H_{\rm ZP}(z)$, which results from inserting $f$ zeros in between the individual coefficients of ${\mathbf h}$. Furthermore, let~ $z_{0,\,1},...,z_{0,\,G}$ denote the zeros of $H(z)$. An insertion of $f$ zeros in the time domain corresponds to a transform $z \mapsto z^{1/(f+1)}$ in the $z$-domain, i.e., $H_{\rm ZP}(z)\!=H(z^{f+1})$. This means, the $(f+1)G$ zeros of $H_{\rm ZP}(z)$ are given by the $(f\!+\!1)$ complex roots of $z_{0,\,1},...,z_{0,\,G}$, respectively. Consider a certain zero~ $z_{0,\,g}\!:=\!r_{0,\,g}\,{\rm exp}({\rm j}\,\varphi_{0,\,g})$~ of $H(z)$ that is outside the unit circle ($r_{0,\,g}\!>\!1$). This zero will lead to $(f\!+\!1)$ zeros
\begin{equation}
z_{0,\,g}^{(\lambda)}\;:=\;r_{0,\,g}^{1/(f+1)}\,{\rm exp}\left(\;{\rm j}\,\frac{2\pi\lambda + \varphi_{0,\,g}}{f+1}\;\right)
\end{equation}
of $H_{\rm ZP}(z)$ ($\lambda = 0,...,f$) that are located on a circle of radius~ $r_{0,\,g}^{1/(f+1)}\!>\!1$, i.e., also outside the unit circle. By means of minimum-phase prefiltering, these zeros are reflected into the unit circle, i.e., the corresponding zeros of $H_{\rm ZP,min}(z)$ are given by $1/z_{0,\,g}^{(\lambda)*}$. Therefore, the sparse CIR structure is retained after minimum-phase prefiltering (with the same zero-pad grid), since the zeros of $H_{\rm ZP,min}(z)$ are the $(f\!+\!1)$ roots of the zeros of $H_{\rm min}(z)$. Specifically, the non-zero coefficients of ${\mathbf h}_{\rm ZP,min}$ are given by the CIR ${\mathbf h}_{\rm min}$. If the zeros of $H(z)$ (or equivalently of $H_{\rm ZP,min}(z)$) are not too close to the unit circle, ${\mathbf h}_{\rm min}$ is characterized by a significant energy concentration in the first channel coefficients. In this case the effective channel memory length of ${\mathbf h}_{\rm ZP}$ is significantly reduced by minimum-phase prefiltering, namely by some multiples of $(f\!+\!1)$, cf.~(\ref{sparseCIR}). 

\subsection{Numerical Results}
In the sequel, numerical results obtained by Monte-Carlo simulations are presented. To start with, a static sparse ISI channel is considered, and the BER performance of the proposed receiver structure is compared with the $\mbox{sub-P-BCJRA}$ equalizer \cite{parallelMAP}. As an example, we consider here the CIR \mbox{${\bf h}=[\, h_0 \; 0\hdots 0 \; h_4 \; 0\hdots 0 \;h_7\; 0\hdots 0 \; h_{15}\,]^{\rm T}$} with $h_0\!=\!0.87$ and $h_4\!=\!h_7\!=\!h_{15}\!=\!0.29$ from \cite{parallelMAP}, which has a general sparse structure (i.e., no zero-pad structure). The BER performance (binary antipodal transmission, $M\!=\!2$) of the $\mbox{sub-P-BCJRA}$ equalizer and the DDFSE equalizer with WMF is displayed in Fig.~3, as a function of $E_{\rm b}/N_0$ in dB, where $E_{\rm b}$ denotes the average energy per bit and $N_0$ the single-sided noise power density ($E_{\rm b}/N_0\!:=\!1/\sigma_{\rm n}^2$). Due to the given channel memory length, the complexity of MLSE detection is prohibitive. As a reference curve, however, the matched filter bound (MFB) is included, which constitutes a lower bound on the BER of MLSE detection. The filter length of the WMF has been chosen as $L_{\rm F}\!=\!40$. Since the channel is static, the filter has to be computed only once. When the parameter $K$ is chosen as $K\!=\!4$, the overall receiver complexity is approximately the same as for the $\mbox{sub-P-BCJRA}$ equalizer. As can be seen in Fig.~3, the BER performance achieved by the proposed receiver structure is also comparable to the $\mbox{sub-P-BCJRA}$ equalizer. At a BER of $10^{-3}$, the loss with respect to the MFB is only about 1~dB. At the expense of a small loss due to residual ISI ($0.5$~dB at the same BER), the complexity of the DDFSE equalizer can be further reduced to $K\!=\!3$. 

\begin{figure}[t]
\begin{center}
\includegraphics*[scale=0.47]{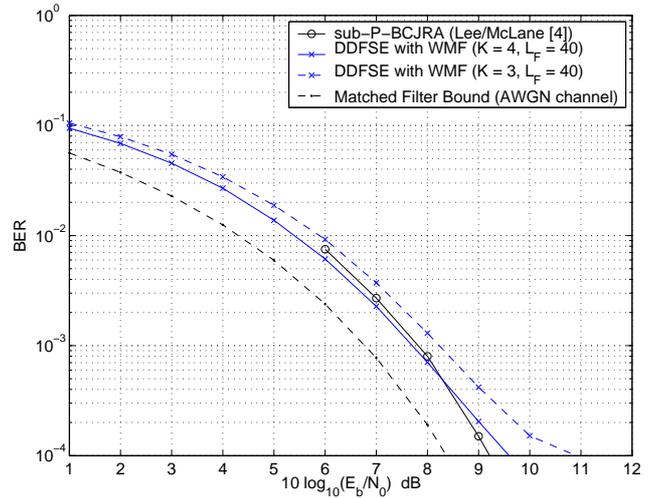} 
\end{center}\vspace*{-0.2cm}
\caption{BER performance of the proposed receiver structure in the case of a static sparse ISI channel.}
\end{figure}

Next, we consider the case of a sparse Rayleigh fading channel model, i.e., the channel coefficients $h_g$ ($g\!=\!0,...,G$) in~(\ref{sparseCIR}) are now zero-mean complex Gaussian random variables with variance ${\rm E}\{|h_g|^2\}\!=:\!\sigma_{{\rm h},g}^2$. It is assumed in the sequel that the individual channel coefficients are statistically independent. Moreover, block fading is considered for simplicity. As an example, we assume a CIR with $G\!=\!3$ and a power profile
\begin{equation}
{\bf p}\;:=\;[\,\sigma_{{\rm h},0}^2 \; \underbrace{0\hdots 0}_{ f \; {\rm zeros}} \; \sigma_{{\rm h},1}^2 \;\;0\;\;0\;\;0\;\;\;\sigma_{{\rm h},2}^2 \;\sigma_{{\rm h},3}^2\,]^{\rm T}\,.
\end{equation}  
Note that this CIR again does not have a zero-pad structure. By choosing different values for the parameter $f$, different channel memory lengths $L\!=\!f+6$ can be studied. To start with, consider a power profile with equal variances $\sigma_{{\rm h},0}^2\!=\!...\!=\!\sigma_{{\rm h},3}^2\!=\!0.25$. Fig.~4 shows the BER performance of the proposed receiver structure for binary transmission and three different channel memory lengths $L$ (solid lines: $L\!=\!6$, dashed lines: $L\!=\!12$, dotted lines: $L\!=\!20$). The filter length has been chosen as $L_{\rm F}\!=\!20$ ($L\!=\!6$),~ $L_{\rm F}\!=\!36$ ($L\!=\!12$),~ and $L_{\rm F}\!=\!60$ ($L\!=\!20$). As reference curves, the BER for flat Rayleigh fading ($L\!=\!0$) is included as well as the MFB \cite[Ch.~14.5]{proakis}. In the case $L\!=\!6$, MLSE detection is still feasible. As can be seen in Fig.~4, its performance is very close to the MFB. The DDFSE equalizer with $K\!=\!5$ in conjunction with the WMF achieves a BER performance very close to MLSE detection (the loss at a BER of $10^{-3}$ is only about $0.6$~dB). Even when the channel memory length is increased to $L\!=\!20$, the BER curve of the DDFSE equalizer with WMF deviates only 2~dB from the MFB. Specifically, a significant gain compared to flat Rayleigh fading is achieved, i.e., a good portion of the inherent diversity (due to the independently fading channel coefficients) is captured. When the DDFSE equalizer is used without WMF, a significant performance loss occurs already for $L\!=\!6$. For the larger channel memory lengths, the influence of the WMF makes a dramatic difference: The BER increases by several decades when the WMF is not used. 

\begin{figure}[t]
\begin{center}
\includegraphics*[scale=0.60]{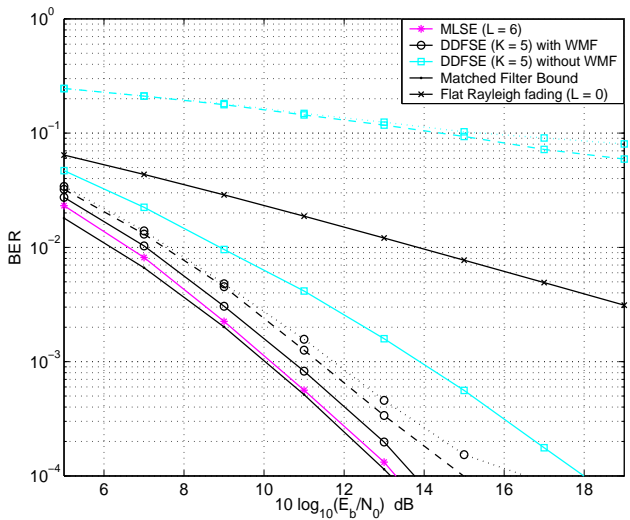} 
\end{center}\vspace*{-0.2cm}
\caption{BER performance of the proposed receiver structure in the case of a sparse Rayleigh fading channel.}\vspace*{-0.3cm}
\end{figure}

Similar performance results have also been obtained for unequal variances $\sigma_{{\rm h},g}^2$. When the power profile ${\bf p}$ of the original CIR already exhibits an energy concentration in the first channel coefficients, the benefit of the WMF is smaller, but still significant. As an alternative to the proposed receiver structure, we have also studied the use of a linear channel shortening filter (CSF)~\cite{ChShort}, which transforms a given CIR with memory length $L$ into a CIR with a reduced memory length $L_{\rm s}\!<\!L$. When $L_{\rm s}$ is chosen small enough, the application of the conventional VA is again feasible, operating on the shortened CIR. However, the performance of this receiver structure turned out to be inferior to the DDFSE equalizer with WMF \cite{Dresden-Paper}.

The concept of minimum-phase prefiltering for sparse ISI channels  is also beneficial when using a tree-based equalization algorithm, such as the LISS algorithm \cite{liss}. In order to obtain a small complexity, the metrics of two competing paths that deviate closely to the root of the tree should differ as much as possible. This is achieved by means of minimum-phase prefiltering, due to the energy concentration in the first channel coefficients of the filtered CIR.

\section{Conclusions}\label{concl}
In this paper, trellis-based equalization of sparse inter\-symbol-interference channels has been revisited. Due to the large memory length of sparse channels, efficient equalization with an acceptable complexity-performance trade-off is a demanding task. With regard to complexity reduction, it has been demonstrated in which cases a decomposition of the conventional trellis diagram into multiple parallel trellises is possible without loss of optimality. In order to tackle general sparse channels, a receiver structure with a linear filter and a reduced-complexity trellis-based equalizer has been studied. The employed equalizer algorithm is standard, i.e., not specifically designed for sparse channels, because the sparse channel structure is normally lost after prefiltering. The coefficients of the linear filter can be computed using standard techniques from the literature. By means of numerical results, the efficiency of the proposed receiver structure has been demonstrated, both for static and fading channels.   

\section*{Acknowledgment}
The authors would like to thank Dr.~Wolfgang Gerstacker (University of Erlangen-Nuremberg, Germany), Ragnar Thobaben (University of Kiel, Germany), and Dr.~J\"org Kliewer (University of Notre Dame, Indiana, USA) for helpful suggestions, especially concerning the prefiltering part.
\vfill
\renewcommand{\baselinestretch}{1}
\bibliographystyle{IEEE}

%

%

\end{document}